\pdfoutput=1
\documentclass[aps,pre,10pt,twocolumn,showpacs,floatfix,nofootinbib,
superscriptaddress]{revtex4-1}
\usepackage{subfigure}
\usepackage{amssymb}
\usepackage{amsfonts}
\usepackage{amsmath}
\usepackage{amsthm}
\usepackage{epsfig}
\usepackage{graphicx}
\usepackage[usenames,dvipsnames]{color}
\usepackage[latin1]{inputenc}
\usepackage{hyperref}
\usepackage{comment}
\usepackage[normalem]{ulem}

\newcommand{\lrangle}[1]{\langle{#1}\rangle}

\begin{document}
\title{Optimal detrended fluctuation analysis as a tool for
the determination of the roughness exponent of the mounded surfaces}

\date{\today}

\author{Edwin E. Mozo Luis}
\email{eluis@ufba.br}
\address{Instituto de F\'{\i}sica, Universidade Federal da Bahia,
   Campus Universit\'{a}rio da Federa\c c\~ao,
   Rua Bar\~{a}o de Jeremoabo s/n,
40170-115, Salvador, BA, Brazil}

\author{Thiago A. de Assis}
\email{thiagoaa@ufba.br}
\address{Instituto de F\'{\i}sica, Universidade Federal da Bahia,
   Campus Universit\'{a}rio da Federa\c c\~ao,
   Rua Bar\~{a}o de Jeremoabo s/n,
40170-115, Salvador, BA, Brazil}

\author{Silvio C. Ferreira}
\email{silviojr@ufv.br}
\affiliation{Departamento de F\'{\i}sica, Universidade Federal de
  Vi\c{c}osa, Minas Gerais, 36570-900, Vi\c{c}osa, Brazil}

\begin{abstract}
We present an optimal detrended fluctuation analysis (DFA) and applied
it to evaluate the local roughness exponent in non-equilibrium surface
growth models with mounded morphology. Our method consists in
analyzing the height fluctuations computing the shortest distance of
each point of the profile to a detrending curved that fits the surface
within the investigated interval. We compare the optimal DFA (ODFA)
with both the standard DFA and nondetrended analysis. We validate the
ODFA method considering a one-dimensional model  in the
Kardar-Parisi-Zhang universality class starting from a mounded initial
condition. We applied the methods to the Clarke-Vvdensky (CV)
model in $2+1$ dimensions with thermally activated surface diffusion
and absence of step barriers. It is expected that this model belongs
to the nonlinear Molecular Beam Epitaxy (nMBE) universality class.
However, an explicit observation of the roughness exponent in
agreement with the nMBE class was still missing. The effective
roughness exponent obtained with ODFA agrees with the value expected
for nMBE class whereas using the other methods it does not. We also
characterized the transient anomalous scaling of the CV model and
obtained that the corresponding exponent is in agreement with the
value reported for other nMBE models with weaker corrections to the
scaling.
\end{abstract}

\maketitle

\section{Introduction}

Molecular beam epitaxy (MBE) is a fundamental technique suited to the
production of layered materials driven by vapor
deposition~\cite{Evans}. In particular, crystal quality requires that
each layer is formed before the next one in nonequilibrium surface conditions~\cite{Evans,Krugmound},
which is achieved at high adatom mobility. At sufficiently high
temperatures, the resulting surfaces can be smooth with global
roughness no larger than a few nanometers corresponding to one or two atomic layers.

At moderate temperatures, the interface exhibits kinetic roughening~\cite{Barabasi,Krugmound}.
If the  growth is ruled by surface diffusion, it is expected that the
dynamics in the hydrodynamic limit is described by the non-linear
stochastic equation~\cite{Barabasi}
\begin{equation}
\frac{\partial h}{\partial t} = -\nu \nabla^{4}h + \lambda \nabla^{2} (\nabla h)^{2} + \eta,
\label{Eq1}
\end{equation}
where $h(\textbf{r},t)$ is the height at position $\textbf{r}$ and
time $t$ measured perpendicularly to a $d$-dimensional substrate,
$\eta(\textbf{r},t)$  is a Gaussian, nonconservative noise given by
$\lrangle{\eta}=0$ and $\lrangle{\eta(\mathbf{r},t)\eta(\mathbf{r}',t')}=
D\delta^d(\mathbf{r}-\mathbf{r}')\delta(t-t')$ while $\nu$ and $\lambda$ are
constants. This equation was independently proposed by Villain
\cite{Villain} and by Lai and Das Sarma \cite{LSarma} being also known
as the Villain-Lai-Das Sarma (VLDS) equation and it is a standard
model of the nonlinear molecular beam epitaxy (nMBE) universality
class. This equation has been investigated in the framework of
renormalization
group~\cite{Janssen,Singha,Haselwandter,Vvedensky4,Vvedensky5} and
many features observed in kinetic Monte Carlo simulations have been
elucidated.

%VLDS equation was recently studied for $d=1$ using a renormalization
%scheme without rescaling at one-loop order and renormalized second and
%third moments were calculated at large-scale and long-time limits
%\cite{Singha}. In the other hand,
%the coefficients of Langevin
%equation (with VLDS form) representing random deposition and surface
%diffusion of adatoms, has been determined for 2-dimensional case,
%providing initial conditions for renormalization group transformations
%\cite{Haselwandter}. This allows to capture the main features from
%Kinetic Monte Carlo (KMC) simulations.

MBE is often modeled by microscopic transition rules built to capture
the atomistic mechanisms. The basic approach is to use stochastic
transition rates for atomistic processes such as deposition and
thermally activated adatom hopping~\cite{Vvedensky2,Leal2011a}. A
fundamental example in this class is  the Clarke-Vvedensky
(CV) model \cite{Vvedensky3}, in which deposition occurs at a constant
and uniform rate and the adatom diffusion rate is given by an
Arrhenius law in the form  $D=\nu_0 \exp(-E/k_{B}T)$ where $\nu_0$ is an
attempt frequency, $k_B$ the Boltzmann constant, and $E$ is an energy barrier for the hopping of an
adatom with $n$ nearest neighbors (bonds). The activation barrier
includes the contribution of the substrate ($E_S$) and
bonds in the same layer ($E_N$)  assuming the form $E=E_S+nE_N$.
Renormalization studies~\cite{Haselwandter,Vvedensky4,Vvedensky5}
point out that the CV model belongs to the nMBE class. The presence of step barriers~\cite{Evans} in
the CV model, in
which the diffusion between different atomic layers is depleted, asymptotically leads to mounded morphologies with non
self-affine structure~\cite{Leal2011,Leal2011a}.

Some investigations  of the dynamic scaling of the surface roughness in
CV-type models have been
reported~\cite{Tamborenea,Sarma2,Kotrla,MENG,deAssis2015}. The
interface fluctuation within a window of length $l$ (hereafter
called quadratic local roughness) is defined as
\begin{equation}
\omega_i^2(l,t) ={\lrangle{h^2}_i-\lrangle{h}_i^2},
%\left(\lrangle{h^2}_i-\lrangle{h}_i^2\right)^{1/2},
\label{eq:localrou}
\end{equation}
where $\lrangle{\cdots}_i$ means averages over the window $i$. The
quadratic local interface roughness $\omega^2(l,t)$ is defined as the
average of $\omega_i^2$ over different windows and samples. In self-affine dynamical scaling, the local roughness
increases as $\omega\sim t^\beta$ for $t\ll l^z$ and saturates as
$\omega\sim l^\alpha$ for $t\gg l^z$, with $z=\alpha/\beta$, and
these scaling exponents are called growth ($\beta$), roughness
($\alpha$) and dynamic ($z$) exponents~\cite{Barabasi}, respectively.
If $l$ is chosen to be the system size $L$, Eq.~\eqref{eq:localrou}
yields the squared global roughness. Former works on the CV model suggest
temperature dependent exponents and
anomalous scaling of the surface roughness~\cite{Tamborenea,Sarma2,
Kotrla}. Recently, the local roughness of the CV model in $d=2$ has been analyzed and a transient anomalous (non self-affine) scaling has been
found~\cite{deAssis2015}. As a consequence,  nMBE asymptotical
exponents with large corrections to the scaling were conjectured for the CV model,
in consonance with other nMBE models~\cite{Reis2013}.
Numerical simulations in $d=2$ of atomistic models in the nMBE class
with weaker corrections to the scaling provide  growth and dynamical
exponents $\beta\approx0.2$ and $z\approx3.33$,
respectively~\cite{Reis2004} in agreement with the one-loop
renormalization group exponents $\beta=1/5$ and $z=10/3$ for the VLDS
equation \cite{Villain,LSarma}. These exponents were also found from
the scaling analysis of the CV model in Ref.~\cite{deAssis2015}.

However, the scaling of the local roughness at short scales
does not provide estimates of the roughness exponent in agreement with the universality classes they
belong for several irreversible growth models in both one- and two-dimensional
substrates~\cite{Chame}. This caveat was also observed for the CV model in
both irreversible \cite{deAssis2013} and reversible \cite{deAssis2015}
aggregation versions. Then, a direct measurement of the roughness
exponent corresponding to CV model is still lacking.

In the present work, we propose a modification of the standard
detrended fluctuation analysis (DFA)~\cite{Peng}, in which the height
fluctuations are reckoned in terms of the shortest distance to  a
detrending curve, as a method
to calculate the roughness exponent of mounded surfaces.  The method was
validated with a model in the Kardar-Parisi-Zhang (KPZ)~\cite{KPZ}
universality class growing from mounded initial conditions and applied
to the CV model in the regime exhibiting mounded morphologies. Our
method was able to capture fluctuations within length scales smaller
than the mound sizes, arising as a promising strategy to unveil the
universality class of models and experiments where the most relevant contributions to the
fluctuations are within the mounds. This method also provides evidences of
transient anomalous scaling in the CV model with the same
characteristic exponent observed for other nMBE models with weaker
corrections to the scaling.

The sequence of the paper is organized as follows. In
Sec.~\ref{sec:methods}, we present basic concepts and definitions of
kinetic roughening theory and introduce the optimal DFA (ODFA) method.
In Sec.~\ref{sec:KPZ}, we validate the method using the etching
model~\cite{Melo}, that belongs to the  KPZ universality class,  for a
mounded initial condition. In Section~\ref{sec:CV}, we discuss the
formation of rough mounds in CV model and the outcomes obtained with ODFA
method comparing them with DFA and nondetrended methods.
Section~\ref{sec:conclu} summarizes our conclusions.

\section{Methods}

\label{sec:methods}

\subsection{Dynamical scaling}
\label{sec:dynsca}

We will consider the surface evolution in the growth
regime where the global roughness scale as $W=\omega(L,t)\sim
t^\beta$. A characteristic surface length can be extracted from the
autocorrelation function defined as
\begin{equation}
\Gamma(\mathbf{r},t) = \lrangle{\tilde{h}(\mathbf{r_{0}}+\mathbf{r},t)\tilde{h}(\mathbf{r},t)},
\label{Eq5}
\end{equation}
where $\tilde{h}= h - \bar{h}$, $\bar{h}$ is the mean height of
profile, and averages in Eq.~\eqref{Eq5} are performed over different
reference positions $\mathbf{r_{0}}$, different orientations and
independent samples. Surface growth under nonequilibrium conditions
may present mounded
morphologies~\cite{Evansmound,Krugmound,Leal2011a}. In mounded
surfaces, the characteristic lateral surface length can be estimated
as the first zero ($\xi_{0}$) or the first minimum ($\xi_{m}$) of
autocorrelation function~\cite{Evans,Leal2011,Leal2011a}. Those
lengths are expected to scale as
\begin{equation}
\xi_{0,m} \sim  t^{1/z_c},
\label{Eq6}
\end{equation}
where $z_c$ is the coarsening exponent that, in case of self-affine growth, corresponds to the dynamical exponent defined previously.

Under the hypothesis of normal (non
anomalous~\cite{Lopez,deAssis2015,Nascimento}) scaling, the local
roughness obeys the Family-Vicsek ansatz~\cite{Family} given by
\begin{equation}
\omega(l,t) \sim t^{\beta}{F}\left(\frac{l}{t^{1/z}}\right),
\label{Eq8}
\end{equation}
where ${F}$ scales as ${F}(x) \sim
x^{\alpha}$ for $x\ll1$ and ${F}(x)= \mathrm{const}$ for $x\gg1$,
leading to $\omega\sim t^\beta$ for $t\ll l^z$ and $\omega\sim
{l}^\alpha$ for  $t\gg l^z$. For anomalous
scaling~\cite{Lopez1997a}, the local roughness follows the modified
ansatz
\begin{equation}
\omega(l,t) \sim t^{\beta}{F_\mathrm{ano}}\left(\frac{l}{t^{1/z}}\right),
\label{Eq8.1}
\end{equation}
where $F_\mathrm{ano}(x)\sim x^{\alpha_\mathrm{loc}}$ if  $x\ll1$ and
$F_\mathrm{ano}(x)=\mathrm{const}$ for $x\gg 1$. Note that if
$\alpha\ne \alpha_\mathrm{loc}$, one has $\omega(l,t)\sim
l^{\alpha_\mathrm{loc}}t^\kappa$, where
$\kappa=(\alpha-\alpha_\mathrm{loc})/z$, for short scales. Therefore,
for anomalous scaling, the amplitude of $\omega$ \textit{vs} $l$ scales as
$t^{\kappa}$, where $\kappa$ is the anomaly
exponent~\cite{Lopez,deAssis2015,Nascimento}.

\subsection{Optimal DFA}
\label{secNDFA}

\begin{figure}[ht]
\centering
\includegraphics[width=0.9\linewidth]{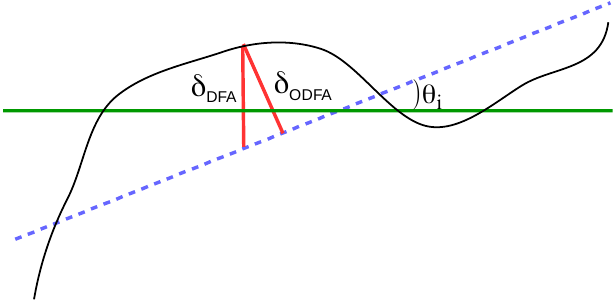}
\caption{Schematic representation of DFA$_1$ and ODFA$_1$ methods. The dashed line represents the linear regression used to detrend the interface and the solid one is parallel to the substrate.}
\label{fig:dfan}
\end{figure}

Let us consider the standard DFA method using a $n$th order polynomial
to detrend the surface~\cite{Peng}, called here of DFA$_{n}$. For sake
of simplicity, we consider one-dimensional cross sections for two-dimensional surfaces. The interface fluctuation
within a window $i$ of size $l$ in DFA$_n$ is defined as
\begin{equation}
\omega^{(n)}_i =
\lrangle{(\delta^{(n)})^2}_i^{1/2}
\label{Eq9}
\end{equation}
where
\begin{equation}
\delta^{(n)}=h(x)-G_i(x;A^{(0)}_i,A^{(1)}_i,\ldots,A^{(n)}_i)
\end{equation}
with $G_i$ being a $n$th order polynomial regression of the interface in
the $i$th window with coefficients
$A^{(0)}_i,A^{(1)}_i,\ldots,A^{(n)}_i$ obtained  using least-square
method~\cite{NR}. The local roughness yielded by the DFA$_n$ method
$\omega^{(n)}$ is defined considering the average over different
windows and samples. We stress that $h(x)$ is the height of the profile with respect to $h=0$. In the standard local roughness
analysis, that corresponds to DFA$_0$, the surfaces fluctuations are
computed in relation to the average height such that
$G_i=A^{(0)}_i=\lrangle{h}_i$.

\begin{figure}[ht]
	\includegraphics [width=8cm]{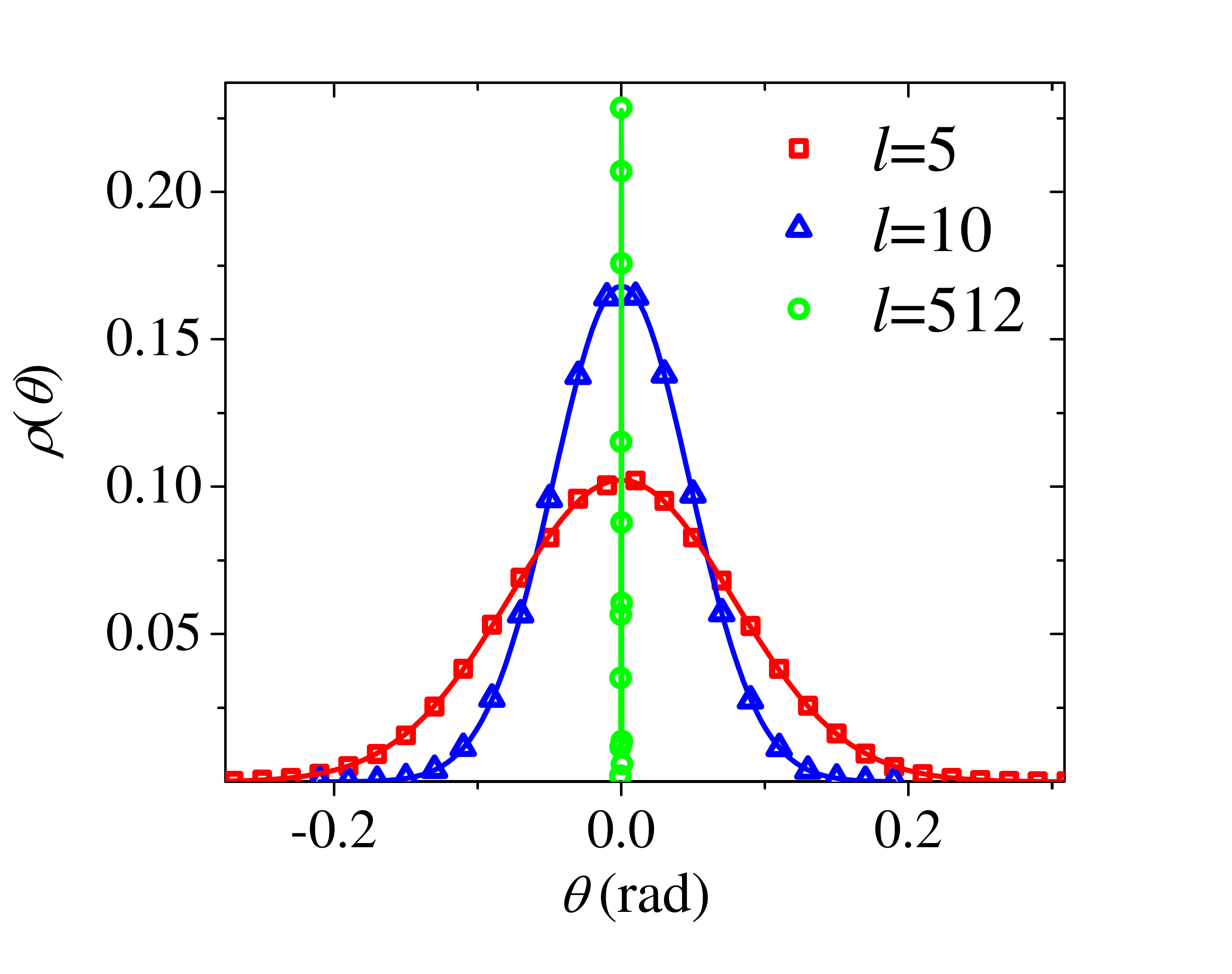}
	\caption{(Color online) Distributions of the local slope for
		different window sizes in a surface obtained with the CV model with $L=1024$, $t=100$, using parameters $R=10$ and $\epsilon=10^{-2}$.}
	\label{Fig0}
\end{figure}
%

%after deposition of $t=100$ monolayers

%
Now, we introduce the ODFA method.
The local roughness in the window $i$ of size $l$ is defined by Eq.~(\ref{Eq9}) with \begin{equation}
\delta^{(n)}=\min\left[h(x)-G_i(x;A^{(0)}_i,A^{(1)}_i,\ldots,A^{(n)}_i)\right], \end{equation} where ``$\min$" represents minimal distance from the surface point with height $h(x)$ to the polynomial $G_i$. ODFA$_0$ corresponds exactly to DFA$_0$ but for higher orders they are different. In particular, we can easily verify that $\delta^{(1)}=[h(x) - {G_{i}}]\cos(\theta_{i})$ is the minimal distance to the detrending curve. Figure~\ref{fig:dfan} shows a schematic representation for DFA$_1$ and ODFA$_1$. The variable $\theta_i$ corresponds to the angle between substrate and the local trend [$\theta_i = \mathrm{atan}(A_i^{(1)})$]. Observe that for $l \gg 1$, both DFA and ODFA correspond to the global roughness given by Eq.~\eqref{eq:localrou} with $l=L$, since the whole surface is not trended. Figure~\ref{Fig0} shows the probability that a window of size $l$ has slope $\theta$ in a simulation of the CV model (see Sec.~\ref{sec:CV} for simulation details). The distributions are Gaussian and converge to Dirac delta functions centered at $\theta=0$ as $l$ increases. In higher order ODFA, the minimal distance can be computed numerically using root finding algorithms as the bisection method~\cite{NR} used in the present work.

\section{Validating the ODFA method}

\label{sec:KPZ}

\begin{figure}[ht]
	\includegraphics[width=0.8\linewidth]{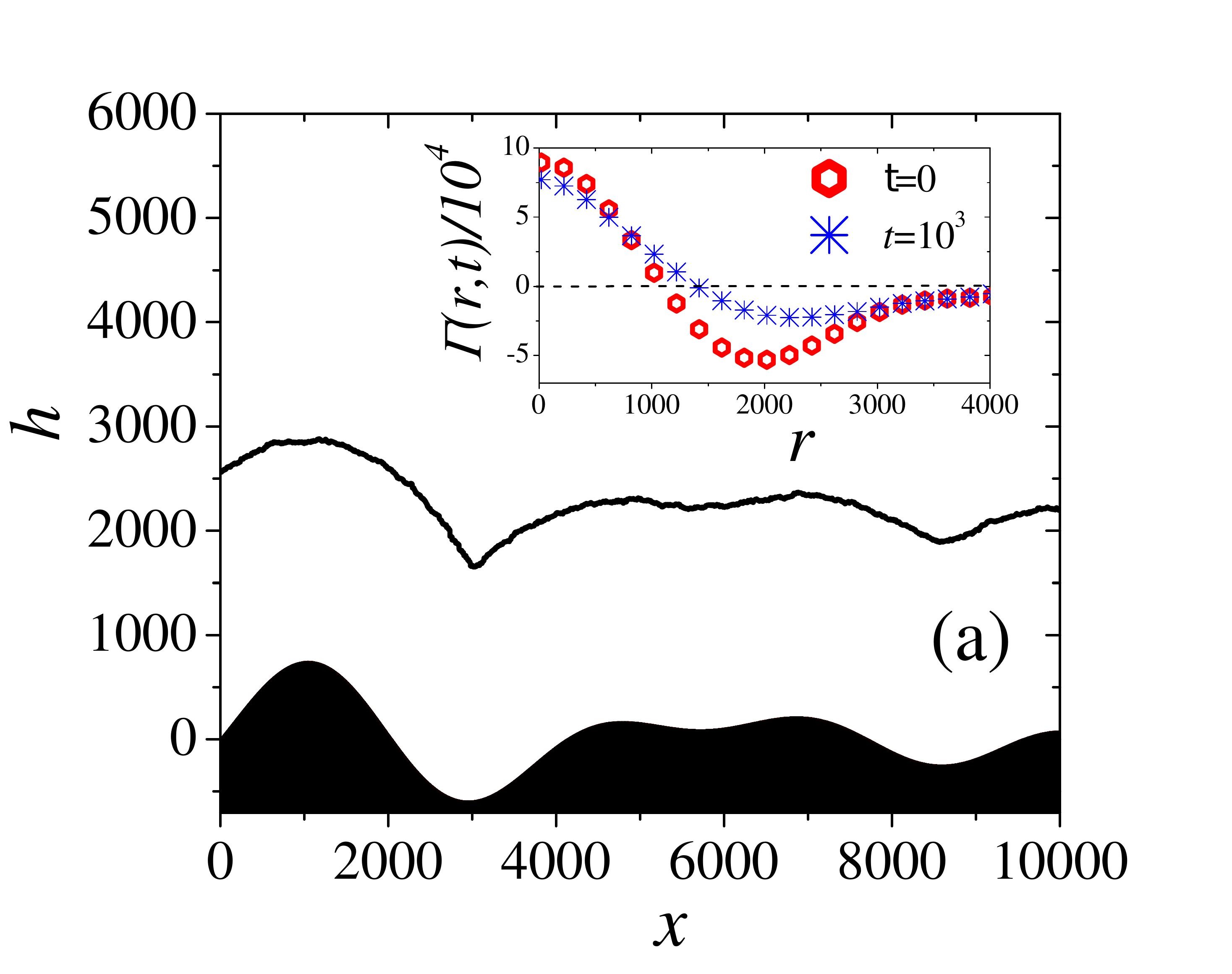}
	\includegraphics[width=0.8\linewidth]{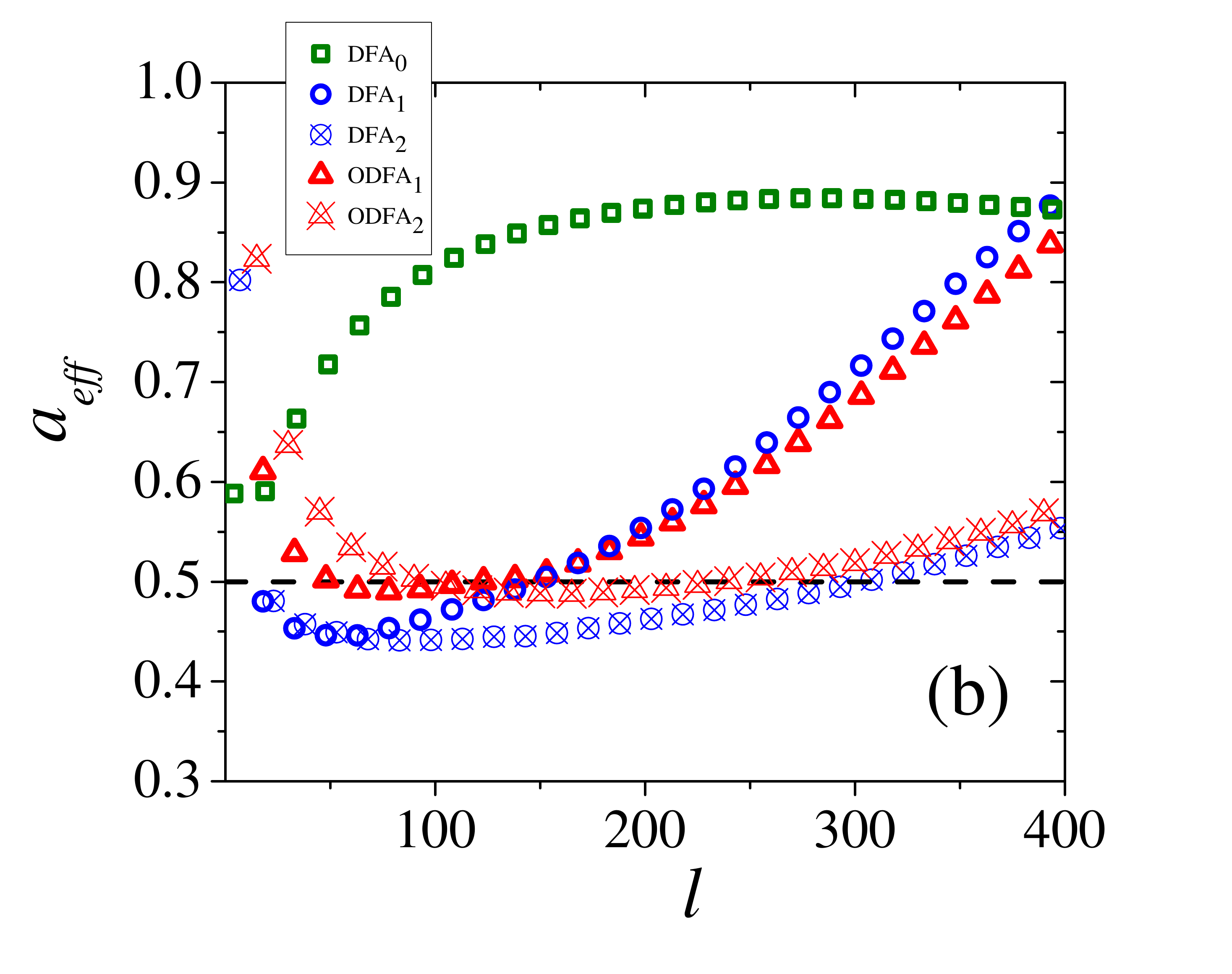}
	\caption{(Color online) (a) Mounded initial condition and the
	(shifted) profile after a deposition time $t=10^3$ using the Etching
	model. The corresponding correlation functions are shown in the
	inset. (b) Effective local roughness exponent as a function of the
	window size for a deposition time $t=10^{3}$ using different methods.
	 The horizontal dashed line indicates the value of the KPZ roughness
	exponent $\alpha_{_\mathrm{KPZ}}=1/2$.}
	\label{Fig4}
\end{figure}

In order to validate the ODFA method and compare it with DFA, we
consider the deposition  on an initially mounded one-dimensional
surface shown in Fig.~\ref{Fig4}(a). The surface evolves according
to the etching model rules~\cite{Melo} in $d=1$ that belongs to
the KPZ universality class~\cite{KPZ}. This model has a large
roughness and also strong corrections to the
scaling~\cite{Oliveira2013,Carrasco2014}, characteristics that make
it suitable to perform the tests. The model rules are as follows: at
each time step, a site $\mathbf{x}$ is randomly chosen and all
nearest-neighbors of $\mathbf{x}$ that obey
$h(\mathbf{y})<h(\mathbf{x})$ are increased as
$h(\mathbf{y})\rightarrow h(\mathbf{x})$ and subsequently
$h(\mathbf{x})\rightarrow h(\mathbf{x})+1$. The surface
preserves its initially mounded structure ($\xi_{0} \simeq 1000$) in the
whole interval of simulation considered, as shown by the correlation
function in the inset of Fig.~\ref{Fig4}(a).

Figure~\ref{Fig4}(b) compares the effective exponent analyses for the local
interface roughness using non detrended analysis (DFA$_0$), DFA and
ODFA up to second order. The effective roughness exponent
$\alpha_\mathrm{eff}$ is defined as $\alpha_\mathrm{eff} \equiv d \ln
(\omega^{(n)}) /d\ln (l) $. DFA$_0$ leads to a large exponent, typical
of mounded surfaces because it is dominated by the long wavelength scales
and does not capture the local fluctuations. The DFA$_1$ analysis provides
a plateau at $\alpha_\mathrm{eff}=0.44$ below the exactly known value
$\alpha_{_\mathrm{KPZ}}=1/2$~\cite{KPZ}.  Using DFA$_2$, we observe an
increasing of the plateau but no significant improvement of
the exponent value is found. For ODFA method however, we observe a
plateau at $\alpha=0.49$ very close to the KPZ exponent $1/2$ with a larger
plateau for ODFA$_2$. We stress that, in scales larger than the average mound length, the values of the
effective roughness exponent reflect the geometrical aspects of the
mounded surface, \textit{i.e.} $\alpha_\mathrm{eff}>0.5$ is expected.

\section{Results for CV model in $d=2$}

\label{sec:CV}

We performed simulations of the CV model in a simple cubic lattice,
with an initially flat substrate at $h=0$ of lateral size $L$.
Periodic boundaries conditions  along the substrate directions are
considered. Deposition occurs with a flux normal to the substrate
of $F$ atoms per site per unit of time
under a  solid-on-solid condition. The ratio
\begin{equation}
R\equiv\frac{D_0}{F},
\label{Eq2}
\end{equation}
in which $D=D_{0} \epsilon^{n}$ is the hopping rate if an adatom has
$n$ lateral neighbors, is a  control parameter of the model. Here,
$D_{0} = \nu_{0}\exp\left(-E_{s}/k_{B}T\right)$ is the hopping rate of
an adatom with no lateral bond, and $\epsilon =
\exp\left(-E_{N}/k_{B}T\right)$.  One time unit corresponds to the
deposition of $L^2$ adatoms, fixing $F=1$ without loss of generality.
In a deposition event, the atom is adsorbed on the top of the
previously deposited adatom or the substrate site to grant the
solid-on-solid condition. For the same reason, only adatoms at the top
of the columns are mobile and perform random hopping towards the top
of their nearest neighbor sites.

\begin{figure}[ht]
	\includegraphics [width=0.8\linewidth] {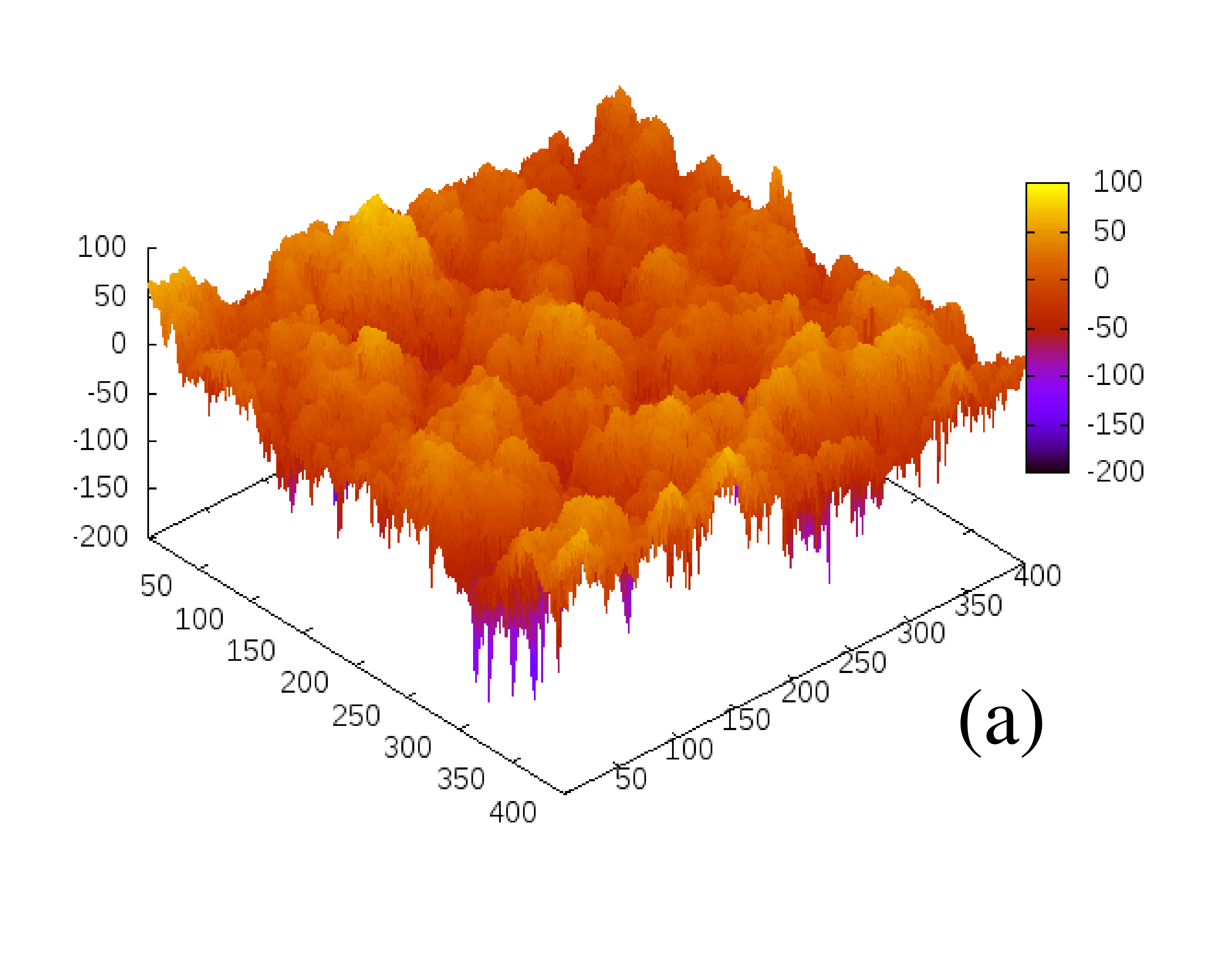}
	\includegraphics [width=0.8\linewidth] {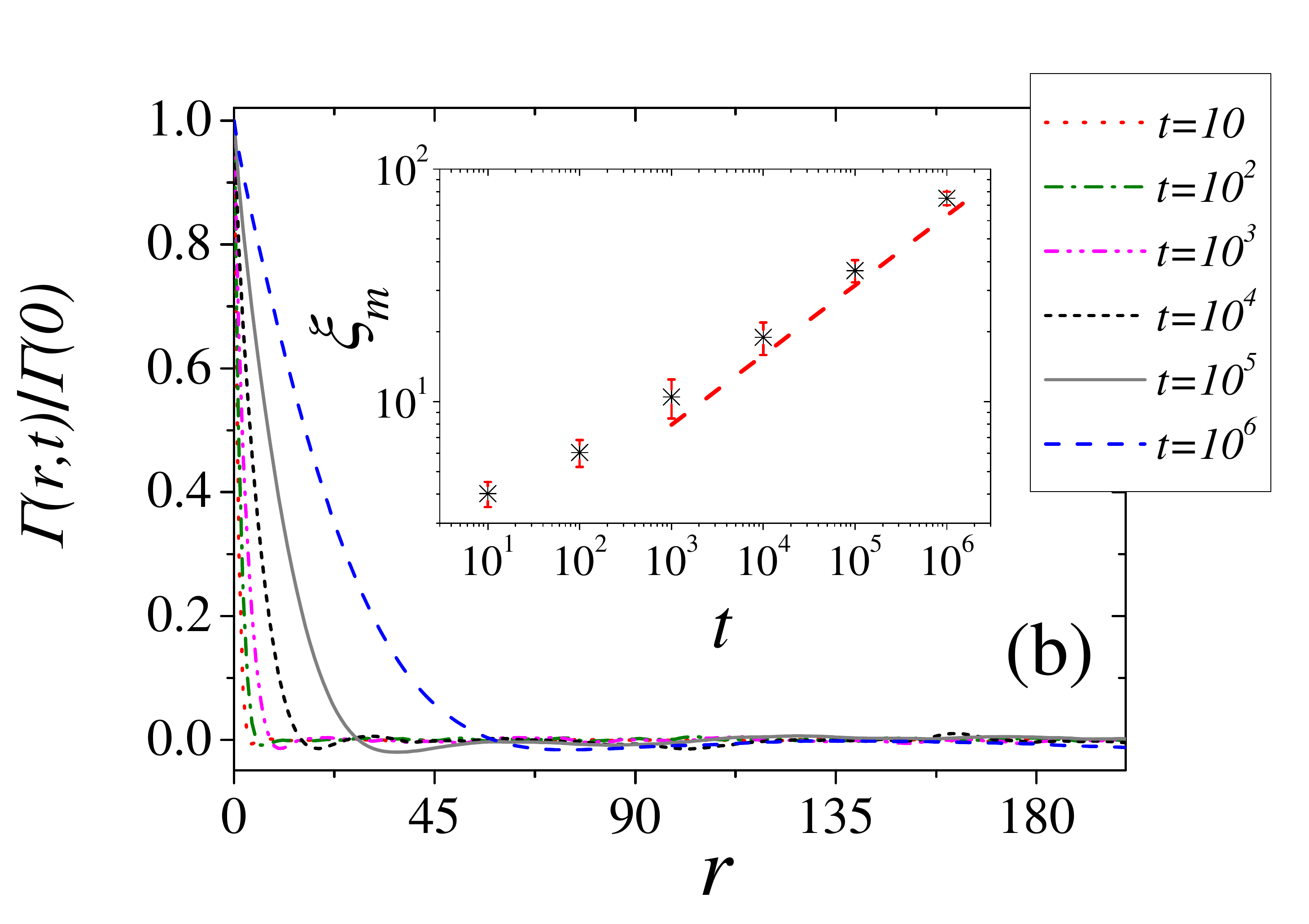}
	\caption{(Color online) (a) Typical morphology of a surface generated
		with the CV model, at deposition time $t=10^{6}$, showing the formation
		of mounds. (b) Scaled autocorrelation function
		$\Gamma(r,t)/\Gamma(0)$ versus distance $r$. The inset
		shows the time evolution of the characteristic length given by the
		minimum of the autocorrelation function. The dashed line has
		a slope 0.30.} \label{Fig1}
\end{figure}

We performed simulations of the CV model for $R=10$ and
$\epsilon=10^{-2}$, which corresponds to a low temperature regime. This
leads to a high surface roughness and allows for a more accurate
scaling analysis since the higher temperature regimes are dominated by
a layer-by-layer growth~\cite{Leal2011,Evans} during an
exceedingly long transient. The substrate has lateral size $L=1024$,
grating that the system is still in the growth regime for the analyzed
times. This parameter set was used in all results presented in this
work but we stress that other values with the same order of magnitude
were studied and the conclusions presented hereafter still holds.

Figure~\ref{Fig1}(a) shows a typical surface morphology after the deposition of $10^{6}$ layers 
generated by simulations of the CV model.
The average size of the mounds was estimated using
the first minimum of the autocorrelation function, as shown in
Fig.~\ref{Fig1}(b). After an initial transient ($t \gtrsim 10^{3}$), a
scaling regime $\xi_{m} \sim t^{1/z_c}$ is observed with an exponent in
agreement with $1/z=0.3$ of the nMBE class as can be seen in the
inset of Fig.~\ref{Fig1}(b), confirming that this lateral
characteristic length has the same scaling as the correlation length
in this regime.

\begin{figure}[ht]
	\includegraphics [width=0.8\linewidth]{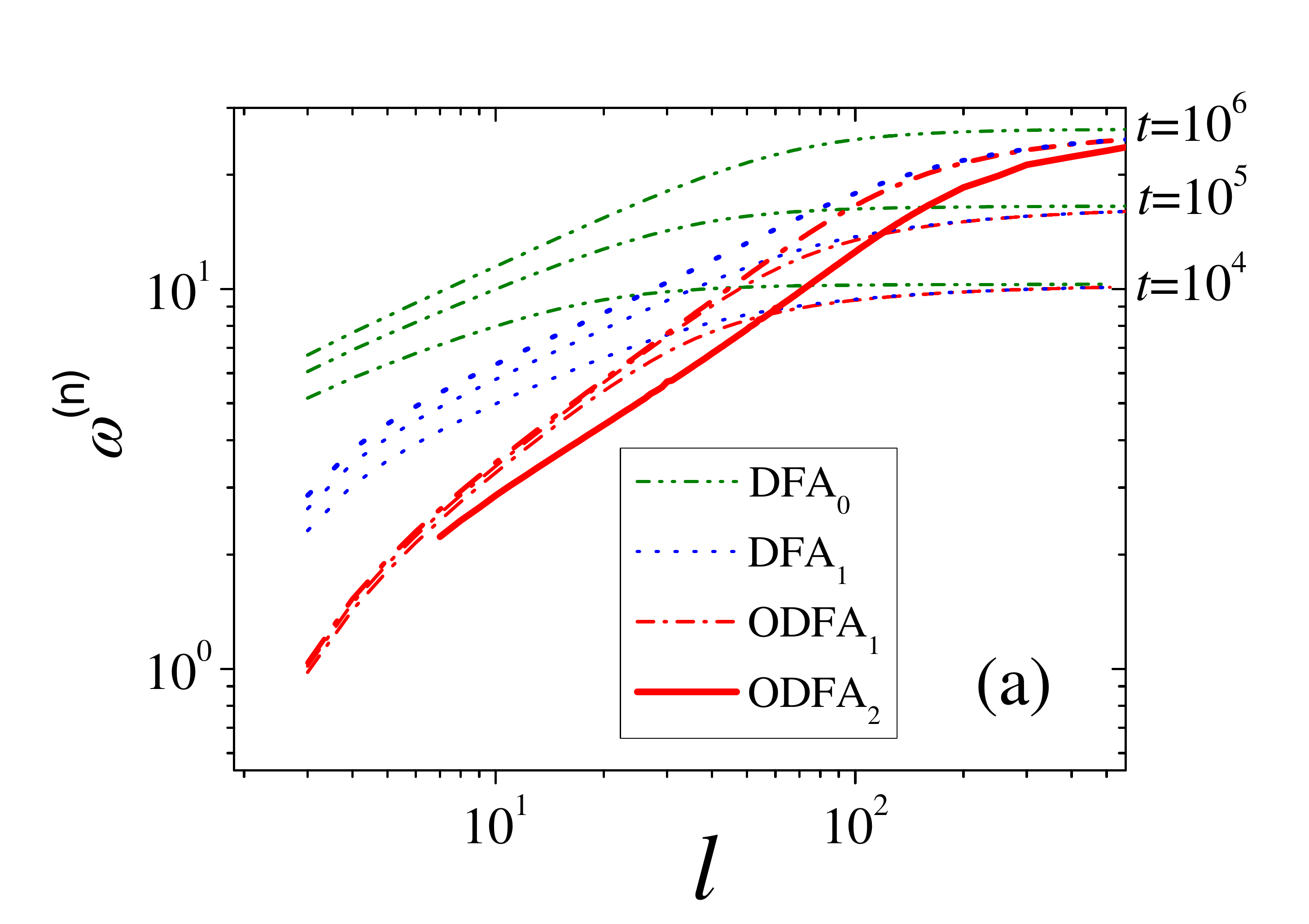}
	\includegraphics [width=0.8\linewidth]{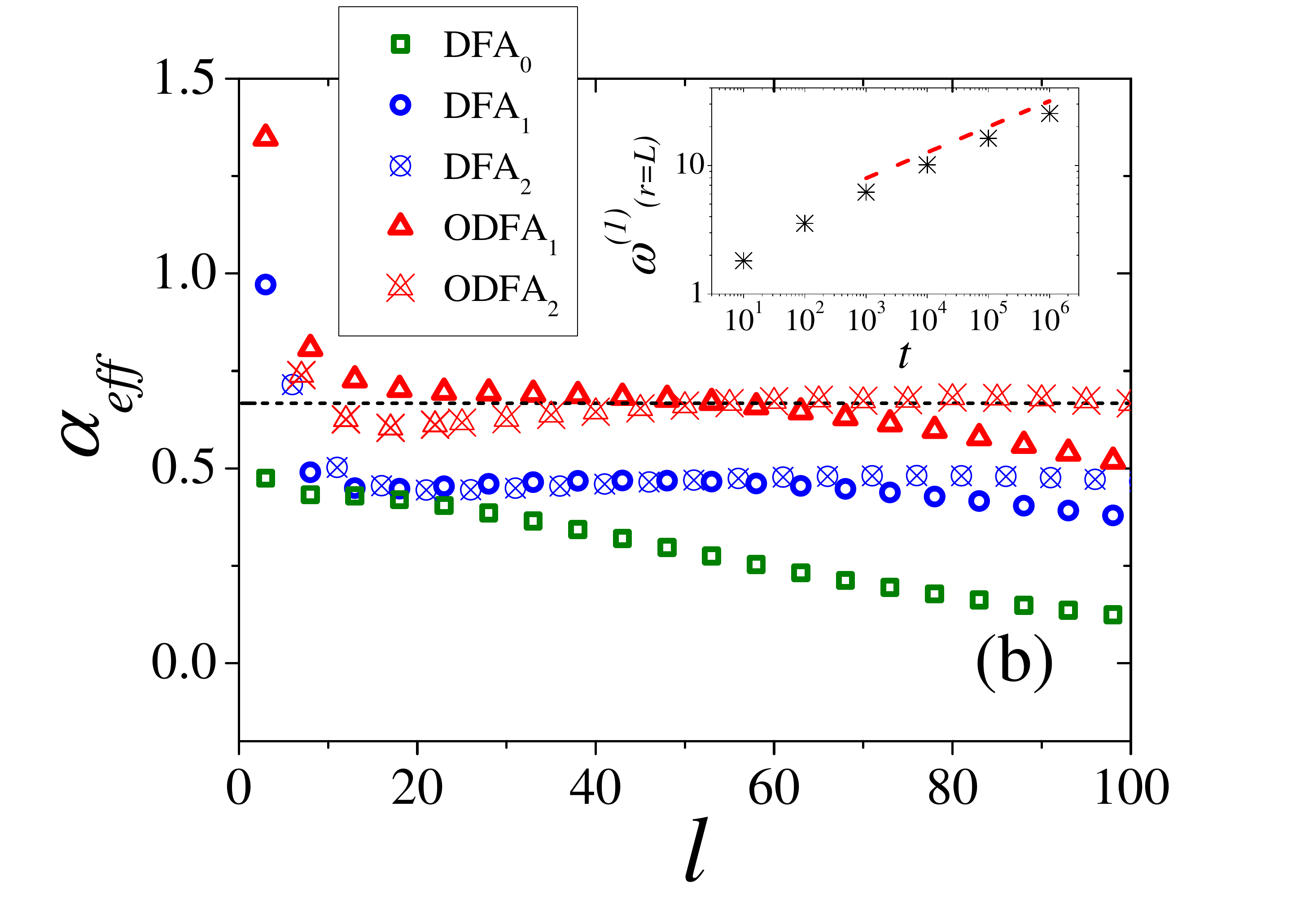}
	\vspace{0.5cm}
	\caption{(Color online) (a) Local roughness using different methods as
	a function of the window size for different times. (b)
	Effective local roughness exponent [$\alpha_\mathrm{eff}\equiv
	d\ln(\omega^{(n)})/d\ln(l)$] analysis for different methods for a deposition
	time $t=10^{6}$. The horizontal dashed line indicates the value of
	the nMBE roughness exponent $\alpha_{_\mathrm{nMBE}}=2/3$. The inset
	shows the time evolution of the global roughness obtained with
	ODFA$_{1}$ and the dashed line has slope 0.20.}
	\label{Fig2}
\end{figure}

Figure~\ref{Fig2}(a) compares the local roughnesses using DFA and ODFA
methods at different times. In Fig.~\ref{Fig2}(b), the effective
roughness exponent is shown as a function of $l$ for $t=10^{6}$.
The analysis for ODFA$_1$ provides a plateau for $\alpha_\mathrm{eff}
\approx 2/3$ for the range $25 \lesssim l \lesssim 60$. A larger plateau is observed for ODFA$_2$. 
We can see, for the time intervals considered, that DFA up second order
provides estimates of the roughness exponent bellow the value expected for 
the nMBE class. Only the sizes of the plateaus are increased for DFA$_2$
similarly to the behavior observed for the etching model in
Fig.~\ref{Fig4}.

From Fig.~\ref{Fig1}(b), we obtain that the average size of the mounds
at $t=10^{6}$ is $\xi_{m} \approx 75$. Therefore, ODFA methods indicate
that nMBE roughness exponent can be extracted from the CV model
considering fluctuations with optimal (minimal) distance to the local
trending within scales up to the same order of the mound size. This result
is particularly useful and raises the possibility of measuring  the
roughness exponent for experimentally accessible times. The
inset of Fig~\ref{Fig2}(b) shows the global roughness against time
computed using  ODFA$_1$ (similar curves were obtained for DFA and ODFA$_2$) for $l =
L$. The results provide $\omega \sim t^{0.2}$ for $t \gtrsim 10^{3}$,
which is fully consistent with the nMBE growth exponent.

\begin{figure}
\includegraphics[width=0.8\linewidth]{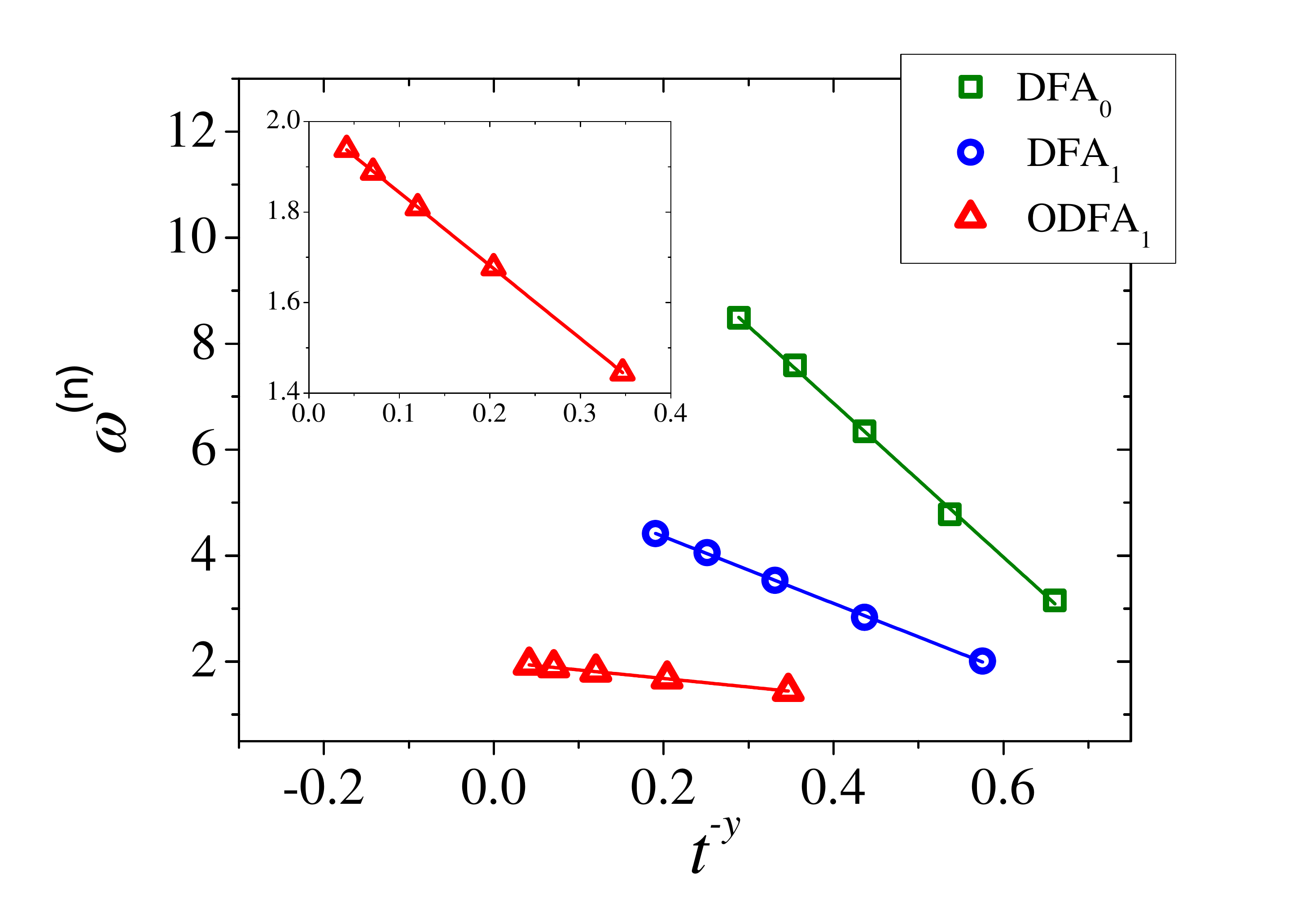}
\caption{(Color online) Transient anomalous scaling analysis for the CV
model using different methods: DFA$_0$ , DFA$_{1}$ and ODFA$_{1}$ with
$y=0.09$, $0.12$, and $0.23$, respectively. The window size used to
compute the local roughness is $l=5$. The inset represents a
zoom of the vertical scale improving the visibility of the ODFA$_1$ curve.}
\label{Fig3}
\end{figure}

Now, we address the transient anomalous scaling of the CV model. In
self-affine (non anomalous) scaling, the local roughness at very
short scales (of order of just a few lattice spaces) approaches a
finite value at long times following a power-law correction given by
\begin{equation}
\omega^{(0)} = C_{1} + C_{2} t^{-y},
\label{Eqlocal}
\end{equation}
where $C_{1}$ and $C_{2}$ are constants. This is the the same behavior
of the average squared local slope $|(\nabla h)|^2$~\cite{Reis2013}.
For the Edwards-Wilkinson (EW)~\cite{EW}  and KPZ~\cite{KPZ}
equations, the values of the exponent $y$ were computed analytically
as $y_{_\mathrm{EW}} = d/2$ and $y_{_\mathrm{KPZ}} = 2 (1 - \alpha)
/z$, respectively~\cite{Krug1,Chou}. In Ref.\cite{deAssis2015}, it was
shown that the effective anomaly exponent in the range $0.08 \leq
\kappa \leq 0.23$ observed for the CV model is due to a transient
effect since the local roughness for a very small scale approaches a
constant value according Eq.~\eqref{Eqlocal} with $y=0.09$.

The local roughness for a window of size $l=5$ is shown as a
function of $t^{-y}$ in Fig.~\ref{Fig3} for DFA$_0$, DFA$_1$ and
ODFA$_1$, considering $y=0.09$, $0.12$ and $0.23$, respectively,
showing that the convergence to a constant value is quicker for
ODFA$_1$. This is an additional evidence that CV model exhibits
asymptotically normal scaling, corroborating conjecture of
Ref.~\cite{Reis2013} where other  models of the nMBE class were
investigated. It is worth noticing that the exponent $y=0.23$ found
with ODFA$_1$  is the same found in the conserved restricted
solid-on-solid (CRSOS) model~\cite{Reis2013}, where scaling
corrections are weaker.

Finally, we notice that ODFA$_n$, with $n\ge 1$,  also  works for self-affine
surfaces and improved results were obtained as compared with DFA$_0$.
However, no significant differences were observed as compared with
DFA$_n$. Therefore, we propose that using  ODFA in experimental and computational
procedures is indicated since it is equivalent to standard
DFA in self-affine growth but captures better the nature of the
fluctuations of mounded surfaces.

\section{Conclusions}

\label{sec:conclu}

In this work, we investigated a detrended fluctuation analysis, in
which the height fluctuations are taken with respect to the optimal
(minimal) distance from the detrending curve. This  modification was
observed to be irrelevant for the determination of the roughness
exponents of purely self-affine surfaces but it
matters for systems with transient mounded behavior,
which encompass models belonging to the important universality class
of non-linear molecular beam epitaxy. The method was validated and
compared with standard DFA analysis using a one-dimensional growth
model with a well-known roughness exponent.

We applied the method to the Clarke-Vvdensky model where deposition
competes with thermally activated surface diffusion producing
interfaces with rough mounds. Since this model possesses the central
mechanisms of the nMBE class, one expects that it exhibits the nMBE
exponents. We compared the present method with non detrended and
standard DFA analyses for the CV model at low temperature and long
times. A roughness exponent in agreement with the nMBE universality
class ($\alpha_{_\mathrm{nMBE}} = 2/3$) was observed only for ODFA.
We also investigated the transient anomalous scaling, in which the
local roughness within small windows converges to a constant value
with a power law correction in time, Eq.~\eqref{Eqlocal}, and found
that the ODFA method yields the same exponent $y=0.23$ observed for
other nMBE models with weaker corrections to the
scaling~\cite{Reis2013}. Since this exponent is universal for other
universality classes, namely EW and KPZ~\cite{Chou,Krug1}, our results
point  that the same holds for the nMBE class.

The reason why  ODFA method is more efficient than DFA is intuitively
simple since the natural distance to a deterministic referential is
the minimal one as illustrated in Fig.~\ref{fig:dfan}. The larger
distances to the detrending curve in  DFA introduces subleading
corrections that are relevant in experimentally
and computationally accessible times and sizes, for which the surfaces actually does
not reach its asymptotic dynamical behavior.

We expect that our results will be of relevance for experimental
analyses, in which mounded morphologies are commonly
observed~\cite{Evans} and the resolution of the universality class is
challenging.  As a perspective, it would be interesting to consider the
role played by the intrinsic smoothening of the surfaces obtained by the widely used scanning probe microscopy techniques that
can markedly interfere in  the roughness exponent
determination~\cite{Lechenault2010} or even  indicate a
misleading universality class~\cite{Alves2016c}.

\section*{Acknowledgements}

The authors acknowledge the financial support of the Conselho Nacional
de Desenvolvimento Cient\'{i}fico e Tecnol\'{o}gico  (CNPq). SCF thanks the
financial support of Funda\c{c}\~{a}o de Amparo \`{a} Pesquisa do Estado de Minas
Gerais (FAPEMIG). EEML acknowledges the support by Coordena\c{c}\~{a}o
de Aperfei\c{c}oamento de Pessoal de N\'{i}vel Superior (CAPES). TAdA
thanks F. D. A. Aar\~{a}o Reis and J. G. V. Miranda for fruitful discussions.

%\vspace{1.0cm}

%\bibliography{dfan5}


%

\end{document}